\documentclass[5p]{elsarticle} 

\usepackage{mathtools}
\usepackage{subcaption}
\usepackage{longtable}
\journal{ }

\usepackage{placeins}

\usepackage{eurosym}
\usepackage{threeparttable} 

\usepackage{url}

\usepackage{hyperref}
\hypersetup{colorlinks,allcolors=black}

\usepackage{lineno}
\modulolinenumbers[5]

\usepackage{lineno,hyperref}
\modulolinenumbers[1]
\usepackage{amsmath}
\usepackage{siunitx}
\usepackage{eurosym}
\biboptions{numbers,sort&compress}
\usepackage[europeanresistors,americaninductors]{circuitikz}
\usepackage{adjustbox}
\usepackage{xspace}
\usepackage{caption}
\usepackage{booktabs}
\usepackage{tabularx}
\usepackage{multicol}
\usepackage{float}
\usepackage{graphicx,dblfloatfix}
\usepackage{csvsimple}

\begin{document}
\begin{frontmatter}

\title{Lessons learned from establishing a rooftop photovoltaic system crowdsourced by students and employees at Aarhus University}
\author[mymainaddress]{Marta Victoria}
\author[mymainaddress]{Zhe Zhang}
\author[mymainaddress]{Gorm B. Andresen}
\author[mymainaddress]{Parisa Rahdan}
\author[mymainaddress]{Ebbe K. Gøtske}

\address[mymainaddress]{Department of Mechanical and Production Engineering, iClimate, Aarhus University, Katrinebjergvej 89, 
8200 Aarhus N, Denmark}

\begin{abstract}
Energy communities are promoted in the European legislation as a strategy to enable citizen participation in the energy transition. Solar photovoltaic (PV) systems, due to their distributed nature, present an opportunity to create such communities. At Aarhus University (Denmark), we have established an energy community consisting of a 98-kW rooftop solar PV installation, crowdsourced by students and employees of the university. The participants can buy one or several shares of the installation (which is divided into 900 shares), the electricity is consumed by the university, and the shareowners receive some economic compensation every year. The road to establishing this energy community has been rough, and we have gathered many lessons. In this manuscript, we present the 10 largest challenges which might arise when setting up a university energy community and our particular approach to facing them. Sharing these learnings might pave the way for those willing to establish their own energy community. We also include policy recommendations at the European, national, and municipality levels to facilitate the deployment of energy communities.

\end{abstract}
\end{frontmatter}

\section{Introduction} 

The rapid cost reduction experienced by solar photovoltaic (PV) modules in recent years and the possibility of installing PV systems on rooftops close to where electricity is consumed makes distributed PV a key component in future energy systems \cite{Rahdan_2024}. The technology is also ideal for constituting energy communities where citizens co-own the electricity generation facility. Energy communities are described in European legislation as \textit{the strategy to enable collective and citizen-driven energy actions to support the clean energy transition }\cite{Clean_Planet}. However, there is limited practical experience on how to implement energy communities, in particular, those in which universities play an active role. Distributed PV systems have also been shown to be a powerful strategy to supply electricity to low- and mid-income households if adequate policies are in place \cite{Oshaughnessy_2021a, Oshaughnessy_2021b, Fox_2023}. 

\

\begin{figure}[!h]
\centering
\includegraphics[width=0.7\linewidth]{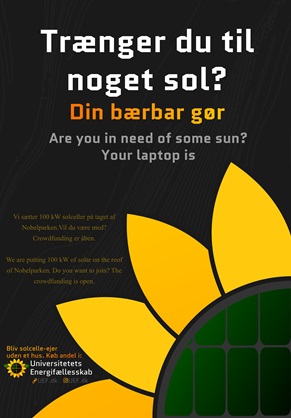}
\caption{Poster used during the campaign to invite students and employees to become part of the solar energy community. Designed by Martin Mørk, one of the members of the UEF energy community.} \label{} 
\end{figure}

From the perspective of our students at the university, two barriers often hinder their potential participation in the energy transition by owning a rooftop PV installation. First, they do not own a house or apartment with available rooftop space. Typically, they live in dormitories or rented apartments that are not perceived as long-term residences. Second, they do not have the initial investment available since, at this time of their life, they are not expected to have a large saving potential. In this context, we have established a crowdsourced rooftop PV installation and invited students and employees of Aarhus University, as well as Aarhus citizens living close by, to buy one or more shares of our common rooftop PV installation. To ensure access to every student, the value of one share corresponds to 109 W and has an initial cost of 900 DKK (120 EUR). The energy community is named “Universitetets Energifælleskab (UEF)” which translates to University Energy Community in Danish. Currently, 120 participants are members of the community, and together they have crowdsourced around 750,000 Danish Krone (100,000 EUR), and installed a 98-kW rooftop solar PV system. All relevant information from our energy community, including the legal statutes that regulate the community, are available through our webpage  \href{www.uef.dk}{www.uef.dk}. 

\

The creation of the energy community at Aarhus University (Denmark) received initial support from the European-funded project AURORA \cite{AURORA_webpage}. Within the framework of this project, similar communities are being developed at the Technical University of Madrid (Spain), University of Evora (Portugal), Ljubljana University (Slovenia) and Forest of Dean District Council (UK).

\

Through the process of establishing the energy community, we have faced several challenges. In this paper, we present the 10 largest challenges that we found, together with the solution or approach that we have taken. Our aim is that the structured collection of these lessons could pave the way for those attempting to establish their own energy community. We also shed light on the required regulatory changes needed to facilitate the implementation of energy communities in Denmark and Europe. 

\section{Challenges and Lessons learned}

In this section, we describe the main questions and challenges that we faced while establishing our energy community, how we addressed them, and what we learned. 

\subsection{Who belongs to the energy community?} \label{sec_who}

When we initially envisioned this project, the university was expected to become a member of the energy community and consume the electricity produced by the PV rooftop installation as a self-consumer. However, we found two challenges, which were common to our colleagues in other European universities attempting to create also an energy community. First, the definition of `renewable energy community' as a legal entity in the European legislation \cite{definition} indicates that \textit{the shareholders can be natural persons, small-medium enterprises, or local authorities, including municipalities}. In many cases, universities are considered national or regionally administrated, and hence it is not guaranteed that the transposition of European legislation in different countries allows universities to become member of an energy community. Second, in Europe, most universities are public institutions subject to legislation that prevents them from getting involved in economic activities that differ from their main purpose (education and research). Getting an exception that allows the university to become a member of the energy community might be legally not possible or require an extremely long time to be approved since it might need the involvement of the university's highest managing body. The problem, if the university is not a member of the energy community and hence not a co-owner of the PV installation, is that in many European countries, self-consumption regulations typically require that the electricity consumer is also the owner of the PV installation. 

\

Our approach at Aarhus University consisted in following an alternative strategy allowed by the existing Danish legislation in self-consumption. A scheme known as `self-consumption via third party' \cite{self_consumption_DK} allows the energy community to be an independent organization that supplies electricity to the university. Together, both organizations set up a supply contract that describes the rules under which the energy community will sell electricity, and the university will purchase it, including the negotiated prices, duration of the contract, etc.  For locally-produced electricity, the “self-consumption via third party” scheme removes the grid tariff but keeps the electricity tax (`\textit{elafgift}' in Danish), Table \ref{tab_components}. The latter represents a much higher cost increment than the former. It should be mentioned that, for normal self-consumers (who own the installation and simultaneously consume the electricity), electricity tax is waived, but not for consumers under "self-consumption via third party". In countries where the legal scheme “self-consumption via third party” does not exist, the university could (i) be part of the energy community to allow using a more classic self-consumption scheme, or (ii) remain independent of the energy community. In the latter case, they can sign a power purchase agreement (PPA) that regulates the conditions under which the energy community and university respectively sell and buy the generated electricity.

In this “self-consumption-via-third-party arrangement”, stakeholders are incentivized in different ways. First, members of the energy community UEF receive annual revenues from the sale of electricity to the university. It is estimated that the investment provides an internal rate of return of approximately 4-5\%. Beyond financial returns, members are also empowered by being actively involved in the decision-making process, fostering a sense of ownership and community engagement. Second, the university enjoys a stable and low-cost electricity supply from the PV installations, while reducing its carbon footprint. Third, although the building owner does not receive direct financial returns from the setup, the learning and practical experiences will help it to make more informed decisions during future implementation of solar PV systems.

\begin{table*}
    \centering
    \begin{tabular}{|ccccc|}
    \hline        
    Components & Paid to &  Self-consumption & Self-consumption  & Reference value \\
    & & & via third party& (EUR cents/kWh) \\
    \hline
    Electricity price   & Electricity deliverer &  & & \\
    \hline
    Transmission tariff     & TSO (Energinet)  & exempt & exempt & 0.82  \\
    \hline
    System tariff      & TSO (Energinet)  & exempt & exempt  & 0.99\\
    \hline
    Net tariff     &  DSO (Konstant)  & exempt & exempt & 1.83 (summer) \\  
      &   &  &  &  
 3.67 (winter) \\ 
    \hline
    Tax (`\textit{elafgift}')     & State & exempt & & 9.66 \\
       \hline
    \end{tabular}
    \caption{Cost component of electricity paid by consumers in Denmark under regulation covering self-consumption (owner and consumer is the same entity) and self-consumption via third party (owner of the installation and consumer of the locally-produced electricity are different entities). Reference values correspond to first quarter of 2025. Net tariff is different for every hour, average values at 06:00-17:00 in Aarhus are reported. TSO stands for Transmission System Operator, which in Denmark is Energinet. DSO stands for Distribution System Operator, which in Aarhus is Konstant.}
    \label{tab_components}
\end{table*}

\subsection{Rules for shareowners} \label{} 

Many practical questions arise when considering the implementation of the community, e.g. How can one person become a member of the energy community? What should be the price for every share? Is there a maximum number of shares per owner? How do we ensure the democratic operation of the community? 

\

\begin{figure}[!h]
\centering
\includegraphics[width=\linewidth]{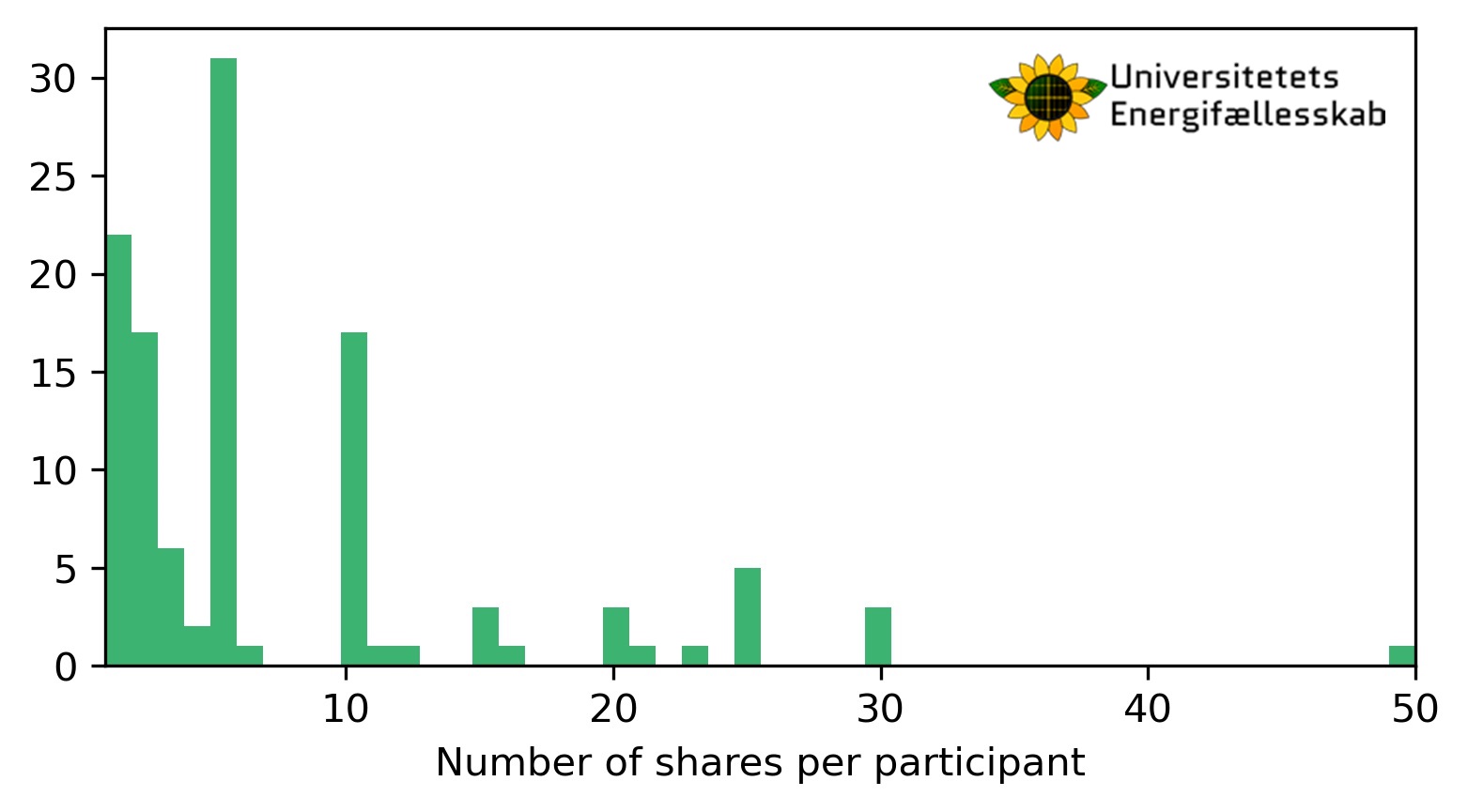}
\caption{Distribution of number of shares per community member at the energy community “Universitetets Energifælleskab (UEF)”. } \label{} 
\end{figure}

In our case, the set of rules describing the ownership of the energy community was decided in a participatory process conducted in several open meetings. These were the basis for the legal statutes of the energy community, which were written with help from our legal advisor and are openly available \cite{statues}. We decided that people who are connected or live close to the university can become part of the energy community, that the community comprises 900 shares, one share is equivalent to 109 W and costs 900 DKK (120 EUR) to ensure accessibility, and that one person can at the maximum own 5\% of the total shares. To ensure democratic governance, we also decided that every participant could cast one vote in our general assembly, regardless of the number of shares that she or he owns. The distribution of the number of shares per member, shown in Fig. 2, indicates that the ownership is quite spread among participants.  As mentioned, these decisions were taken in open meetings with interested participants, while the exact figures for the number of shares and their value were taken once one of the offers to build the installation from the contractor was selected. 

\

The energy community was created in August 2023 by six founding members who took care of creating the first board of the energy community and coordinating the decisions until the first General Assembly took place in May 2024. The members of the board are re-elected every 2 years.

\subsection{Stakeholders management at the university}

One of the most significant challenges experienced by all the universities attempting to create an energy community is related to stakeholder management. This was also the case for our energy community at Aarhus University. The universities are large institutions often lack a clear decision chain when getting involved in innovative proposals, such as the energy community that we presented. For instance, at Aarhus University, stakeholders that were involved in the discussions included, among others, the university director and his team, lawyers at the university, the head of the department that will consume the electricity provided by the energy community, the head of the procurement at the university who is responsible for approving electricity supply contracts, the head of facility management who will need to be involved in some decisions, e.g. where to place the inverters and wires, the climate strategy department at the university, the company owning the university buildings (see Section \ref{FEAS}).

\

Our approach in this case was to first map out the influence-interest matrix of different stakeholders, so we have a better idea of whom we need to manage closely, whose needs we need to meet and who needs to be kept informed without too much engagement. Then, we set up many different meetings to try to explain the rationale of the project to all the different stakeholders and understand their concerns as early as possible. It was also very important to communicate the potential gains that each of them could get from this project. It took us about 2 years to agree with all the relevant stakeholders. This was undoubtedly one of the most effort-consuming tasks of the project, but we must emphasize that this effort is critical. The energy community could only effectively materialise if the relevant stakeholders were on board. Consequently, getting their confirmed approval as early in the process as possible is extremely important. It is also important and effort-demanding to keep all the stakeholders updated and engaged in the steps of the project after the agreement signing, which include the implementation and operation of the PV system. 

\subsection{Establishing the electricity supply agreement at the university}

The agreement to use a constant electricity price and the value of that price was a result of a trade-off. First, we agreed on a constant electricity price. On the one hand, Aarhus University has currently signed an electricity supply contract with a large electricity retailer, with very competitive conditions since the university is a large electricity consumer. This contract specifies that the price of electricity paid by the university fluctuates according to the clearing-market price attained for every hour in the spot market in the country. On the other hand, the energy community would prefer signing an electricity supply contract under a fixed price, which would reduce the risks when calculating the future revenues of the installation. As a trade-off, both parties agreed to sign an electricity supply agreement based on a fixed price which is renewed and can be modified every two years. 

\

Second, negotiating and agreeing on the price at which AU will pay for the electricity produced from UEF’s installation took longer time than we expected. This is due to a few reasons. On one hand, for AU, the price needs to be competitive with what they are paying in the status-quo situation to the existing electricity supplier, to ensure legal compliance and an economic benefit such as savings on electricity bills. On the other hand, the price needs to make a good business case for UEF, considering the additional operating cost (e.g. to remove and reinstall the panels if the roof needs to be renovated, as described in Section \ref{FEAS}). This process required both parties to conduct a thorough investigation of the historical unit price for electricity at AU and its composition, which had remained a black box before this. It also required the energy community to make several estimations on future costs, e.g. the potential cost to remove and reinstall the panels.

\

Unfortunately, the current regulatory framework that covers rooftop photovoltaic generation in Denmark is not favorable for community-based distributed rooftop PV installations. In short, consumers need to pay a tax for locally produced electricity, which is similar to that paid for electricity produced by large power plants and exchanged at the wholesale market. This makes the economic competition very hard, regardless of the additional social benefits brought by self-consumption installation and hinders the development of distributed rooftop PV and energy communities in Denmark. 

\subsection{Agreement to use the rooftop} \label{FEAS}

The identification of the building where the rooftop PV system would be installed should be a priority when starting the project definition. In our case, this step entailed additional complexity due to the fact that the buildings at Aarhus University are owned by a different private institution, see Fig. 3.  We engaged with the building owner early in the project and established an agreement that grants the energy community the right to use the rooftop for the lifetime of the installation. However, this entailed extra costs, the most important being: (i) insurance to cover potential damage to the buildings when constructing the installation and potential liabilities during the operation of the PV systems, and (ii) commitment to remove and reinstall the PV system if the building rooftop requires retrofitting during the following 25 years. Both costs will be covered by the energy community and need to be part of its business plan.

\begin{figure}[!h]
\centering
\includegraphics[width=\linewidth]{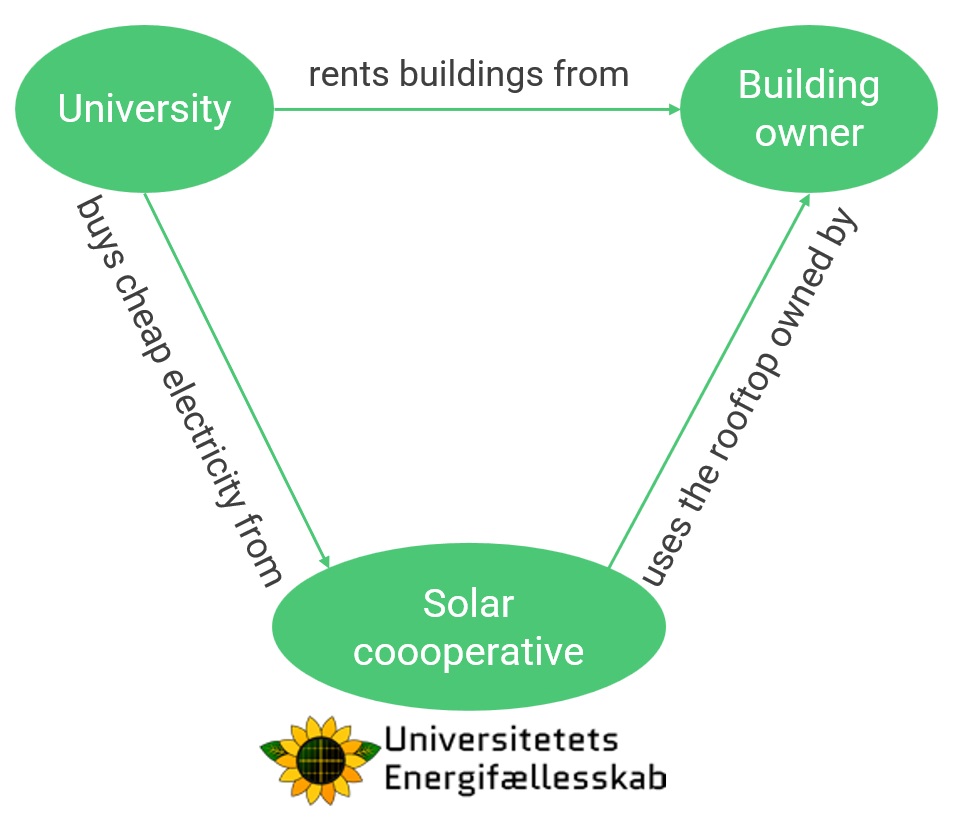}
\caption{Relations between the university (AU), the building owner, and the energy community (solar cooperative).} \label{} 
\end{figure}

\subsection{Organizing the crowdfunding}

To ensure that the energy community has enough funding to finance the rooftop solar PV installation, a crowdfunding campaign was carried out to sell the shares, and a few challenges arose during the process. The first one was to ensure that there was enough interest. In our case, we established a two-step process to buy shares that was very useful. In the first step, participants reserved shares by paying 100 DKK (13 EUR) per share, which allowed us to quantify the real funding capabilities of the community. In the second step, once the contractor has been selected and the final capacity and investment cost of the installation is fixed, the participants pay the remaining 800 DKK (107 EUR) per share. Furthermore, the limit of 5\% shares ownership by a single participant (described in Section \ref{sec_who}), also increased the safety of the process. While we had some participants who reserved shares but did not complete the second step of the payment, in no case this represented a large amount that could put the project at risk. This two-step approach kept participants engaged financially and emotionally, as we observed that some participants increased the number of shares during the second step to help realize the project when they were informed that extra shares became available due to people giving up their reservations.

\

Additionally, the practicality of handling the crowdfunding posed administrative burdens. To the best of our knowledge, there was no Danish online crowdfunding platform that supported the set-up of energy cooperatives due to the complexities of tax reporting. The existing platforms offer functions to collect the funding once, but cannot support the function of distributing yearly dividends back to the crowdfunders. As a result, we adopted a less automatic approach, where bank transfer is used. This led to additional administrative tasks, such as verifying that funding from each participant was received in the bank account of the energy community. 

\subsection{Sizing the PV installation} \label{sec_sizing}

The capacity of the PV installation can be selected based on different objectives. This decision should be taken together with the identification of the building where it would be installed. As a rule of thumb, the design should aim at maximizing the solar electricity that is consumed locally, instead of being exported to the grid. This will improve the business case for the energy community since electricity consumed locally can be charged at a higher price than that obtained when exporting electricity to the grid, because the PV installation typically produces at periods of low spot prices and locally-produced electricity does not pay grid tariffs, see Table \ref{tab_components}. Moreover, exporting electricity to the grid and getting economic compensation might only be possible in some countries. 

\

To maximize self-consumption, one easy first step is trying to identify a building comprising four or more floors. In many cases, this ensures that demand is high so most of the electricity is consumed locally. Additionally, getting access to historic electricity demand time series with hourly resolution to estimate total demand and the hypothetical share of self-consumption for different PV capacities is also recommended. 

\

Of course, the technical sizing of the installation should be done together with setting the target for the funding to be collected in the crowdfunding, as discussed in the previous section. Based on all the previous trade-offs, we selected a target of 100 kW and estimated the self-consumption rate at 85\% using historical demand data. When performing a sensitivity analysis to the internal rate of return of the investment of the energy community members, the self-consumption rate was the assumption with the highest impact on the results, indicating its relevance. Several options exist to maximize self-consumption. First, a battery can be added to the installation, but this option was discarded due to economic reasons. Second, applying demand-side management measurements to increase the temporal overlap of electricity consumption and solar photovoltaic generation. To this end, a screen will be implemented to show all building users real-time solar electricity generation and incentivize them to adapt their consumption patterns. 

\subsection{Estimating Investment and Operation and Management (O\&M) costs}

A description of the investment cost and expenses through the project is key to creating a solid business case for the energy community that ensures that, at the end of the lifetime, the participants have recovered their initial investment and obtained some additional benefits. 

\

\begin{figure}[!h]
\centering
\includegraphics[width=\linewidth]{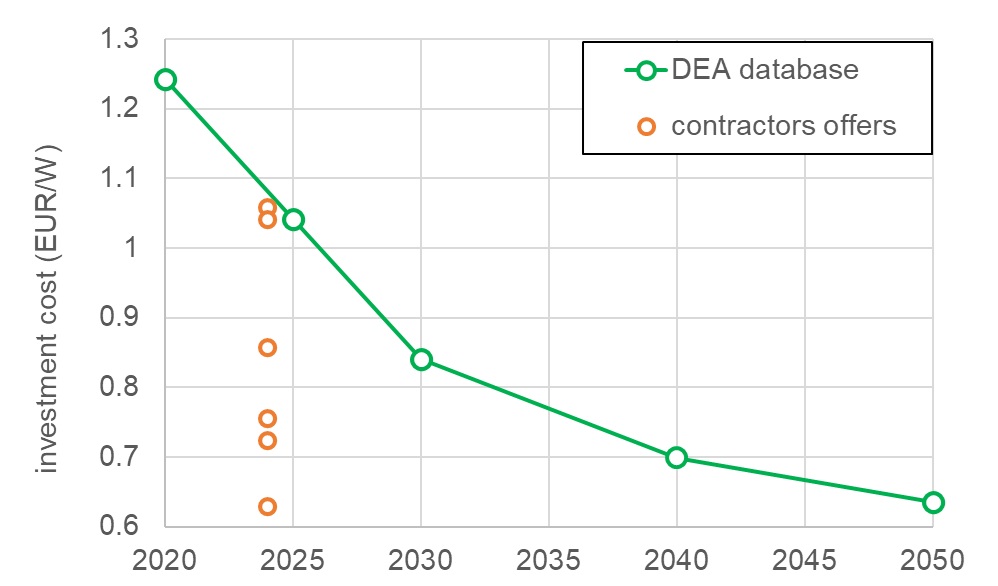}
\caption{Installation costs estimated by the Danish Energy Agency (DEA) for rooftop PV installation \cite{DEA} (green line) and offers received from different contractors in early 2024 for our energy community installation. } \label{fig_investmennt_cost} 
\end{figure}

We use data from the Technology Catalogue by the Danish Energy Agency (DEA) \cite{DEA}, to estimate the investment cost of the installation, see Fig. \ref{fig_investmennt_cost}. Interestingly, the six offers that we got from different contractors were all below the 1.1 EUR/W estimated by DEA, highlighting the continuous cost reduction of distributed PV systems. As it will be described later, we did not select the lowest-cost offer but the one that we found to be more robust in the long term.

While estimating investment cost was relatively easy, performing an estimation of the recurrent cost was more difficult. Besides typical Operation and Management (O\&M) costs for rooftop PV, some additional costs appear in energy communities.  In particular, (a) the cost of insuring the installation during its construction phase as well as during its operation phase; (b) the cost of opening and maintaining a bank account under the name of the energy community; (c) the cost of the administrative work associated with operating the community, including but not limited to, annual reporting to the tax authority, generating invoices to sell the electricity to the university, and distributing the revenues among shareowners; (d) the cost for legal consulting when setting up the community or engineering consulting when evaluating the statics of the building (see Section \ref{sec_implementing}). In our case, costs (a), (b) and (c) are included in the business case of the energy community while costs (d) are supported by the AURORA project.

\subsection{Implementing the PV installation} \label{sec_implementing}

\begin{figure}[!h]
\centering
\includegraphics[width=\linewidth]{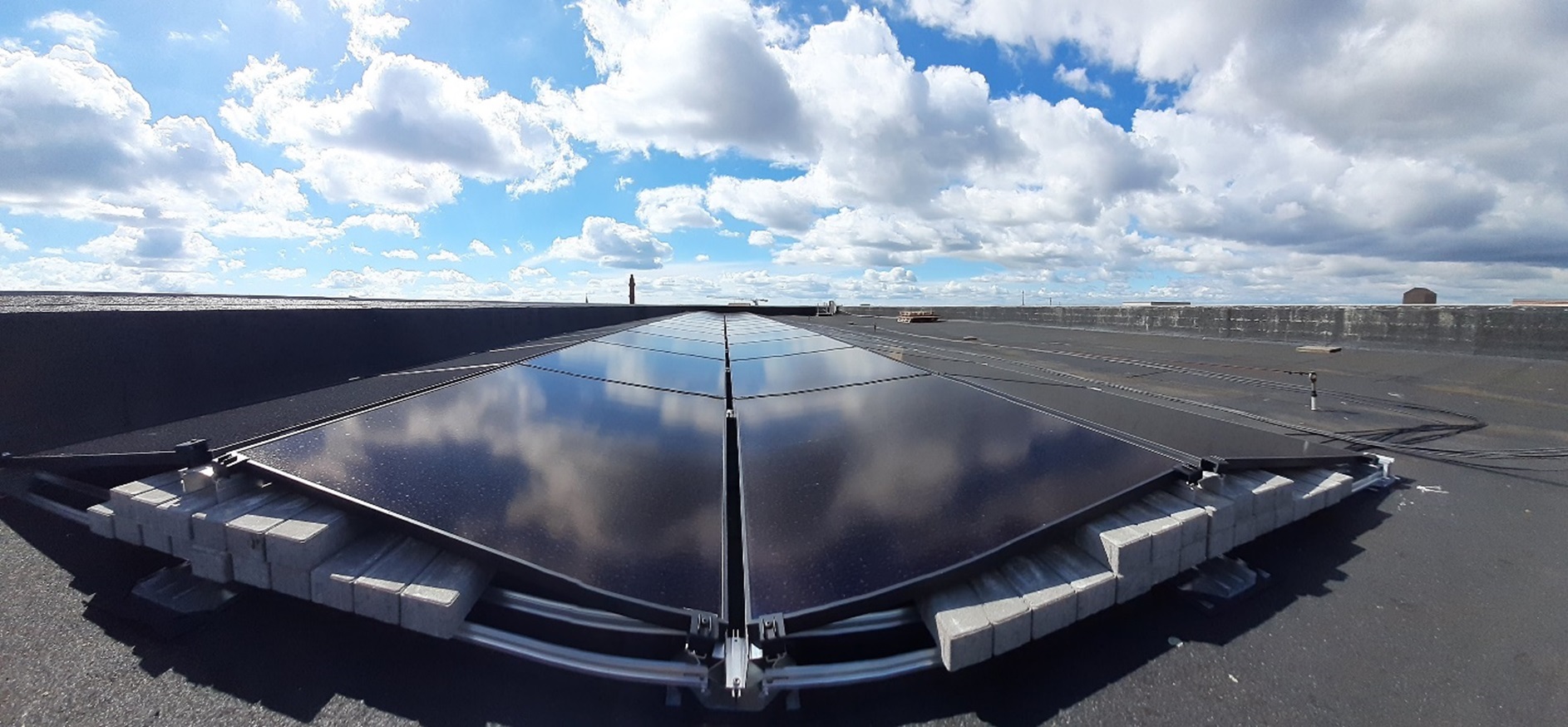}
\caption{Photograph of a section of the installation showing the delta east-west configuration.} \label{fig_photo} 
\end{figure}

One of the critical issues to enable the physical installation of the system is related to the changes in the static load of the building. In our case, the local regulation establishes that changes in the static load of the rooftop above 5\% require a building permit from the municipality, which could have delayed our process for a few months. Hence, our first step was to obtain an engineering evaluation of the maximum load of the PV modules and structures on the rooftop that could be allowed below that threshold. This process was very time-consuming and included (i) consulting the municipality archives to retrieve the documents including a detailed description of the rooftop structure, (ii) contracting a small invasive examination of the rooftop by an engineering consultancy company to verify the documents. 

\begin{figure}[!h]
\centering
\includegraphics[width=\linewidth]{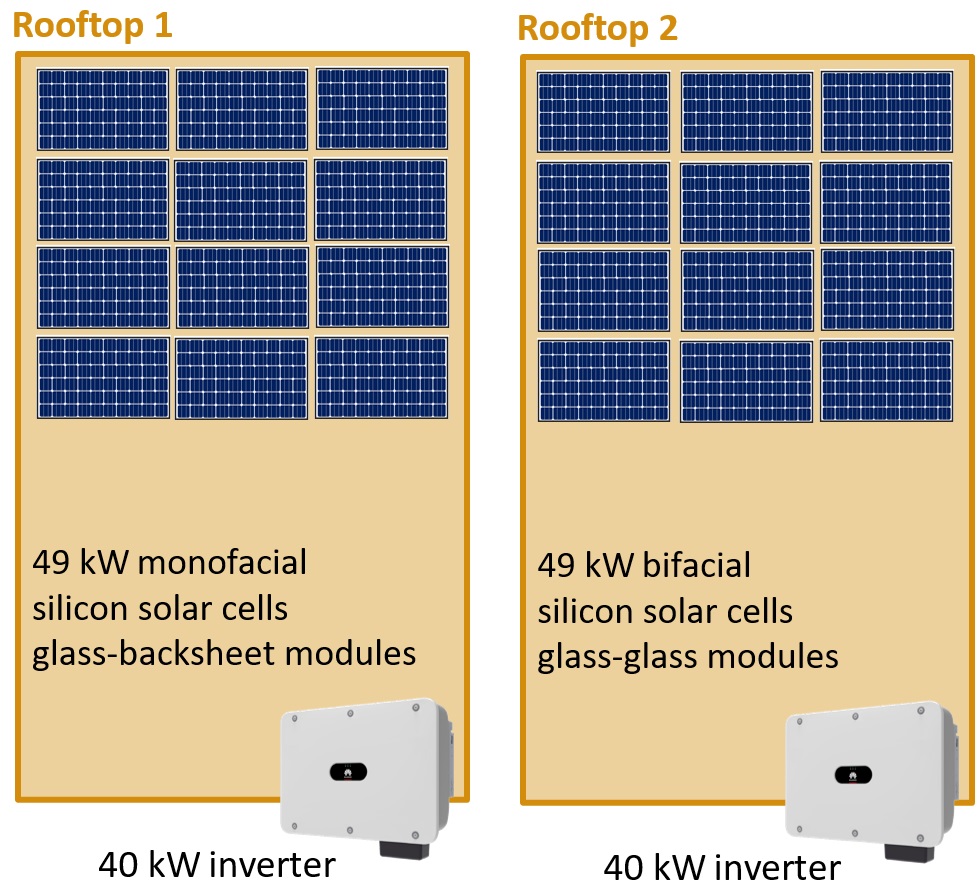}
\caption{Scheme of the PV installations on two different rooftops.} \label{fig_scheme} 
\end{figure}

\

The previous steps allowed us to obtain the maximum weight that could be allowed on the rooftop and the areas on which the PV modules should be placed. This was taken into consideration when selecting among the offers received from different contractors. The board of the energy community assigned the task of evaluating the offer to a small technical committee comprised of members of the energy community. In the evaluation, the criteria for reliability (e.g. selecting a proven configuration) was prioritized over the cost. A delta configuration including two rows of PV modules facing east and west respectively, and tilted 10º was selected (see Fig. \ref{fig_photo}). This was preferred over a classic south-oriented installation for the following reasons. Despite a slightly lower annual generation, the east-west delta configuration reduces the wind loads and, consequently, the need for ballast and weight of the installation. Following the criteria to maximize system reliability over cost, a ballast system was chosen over a glued-structure system because the latter was considered less-proven. Selecting the placement of the inverters also took some discussion. The final agreement was to place them in the building's technical room, where they are protected from outdoor weather. 

\

Since our building is in reality comprised of several towers, the final system includes two subsystems, each of them with its own inverter, see Fig. \ref{fig_scheme}. We took this opportunity to select two types of PV modules (comprising monofacial and bifacial monocrystalline silicon solar cells) and will use our installation to evaluate which of the two performs better under the Danish weather. The installation was completed and the system started producing electricity in October 2024.

\subsection{Operating the installation}

One of the main challenges regarding the operation of the installation is how to ensure the timely completion of all the recurrent activities including (i) checking and invoicing the electricity sold to the university every month, (ii) distributing the economic revenues among the shareowners and (iii) delivering the annual report to the tax authority.  In our case, we have hired an administrator to take care of these activities and included this recurrent cost as part of our business case. 

\

Regarding the selling of electricity to the grid for hours in which the PV generation is higher than the local demand, we are currently evaluating several options. One difficulty is that currently, the electricity market in Denmark allows negative prices, which are typically observed in hours with high renewable generation, e.g. during sunny middays. If we were to sign a contract to sell the excess electricity to the grid, the current Danish regulation establishes that we should also pay when negative prices happen. This poses a risk in our strategy since we could potentially lose money by exporting electricity to the grid and it might be safer to just export the electricity to the grid for free without signing a contract. 
Moreover, most of the companies which could buy our excess electricity require paying certain grid tariffs and handling charges, which reduces the economic benefit of exported electricity. The initial business plan assumed that 85\% of the energy would be locally consumed by the university and the remaining would be exported for free. Hence, whether to export surplus electricity for free and to whom we will sell would also impact our business model and recovery time.

\

We are also currently discussing the procedure to transfer the shares' ownership when some of the participants want to leave the energy community and how to allow new participants in. 

\subsection{Replicating our experience}
\label{sec_replicating}
Our ambition with this project was to become a flagship and show that energy communities can be a reality and include or interact with universities. If we have to name our most important learning, it would be that a very large effort in terms of time and thinking is required to make this a reality. In our case, we could only afford this because we had funding for an employee to push forward this energy community, meet different stakeholders and research alternatives every time we get stuck. Hence, our main suggestion is to make sure that one or several persons have the time to make the energy community happen. We believe replicating this in our local framework would be much easier now that we have collected all the learnings about the process. In this regard, it would be very efficient to have public employees at the municipality level who can enable the deployment of energy communities, by interacting with interested communities, gathering all the necessary information, creating step-by-step guides, etc. 

\begin{figure}[!h]
\centering
\includegraphics[width=\linewidth]{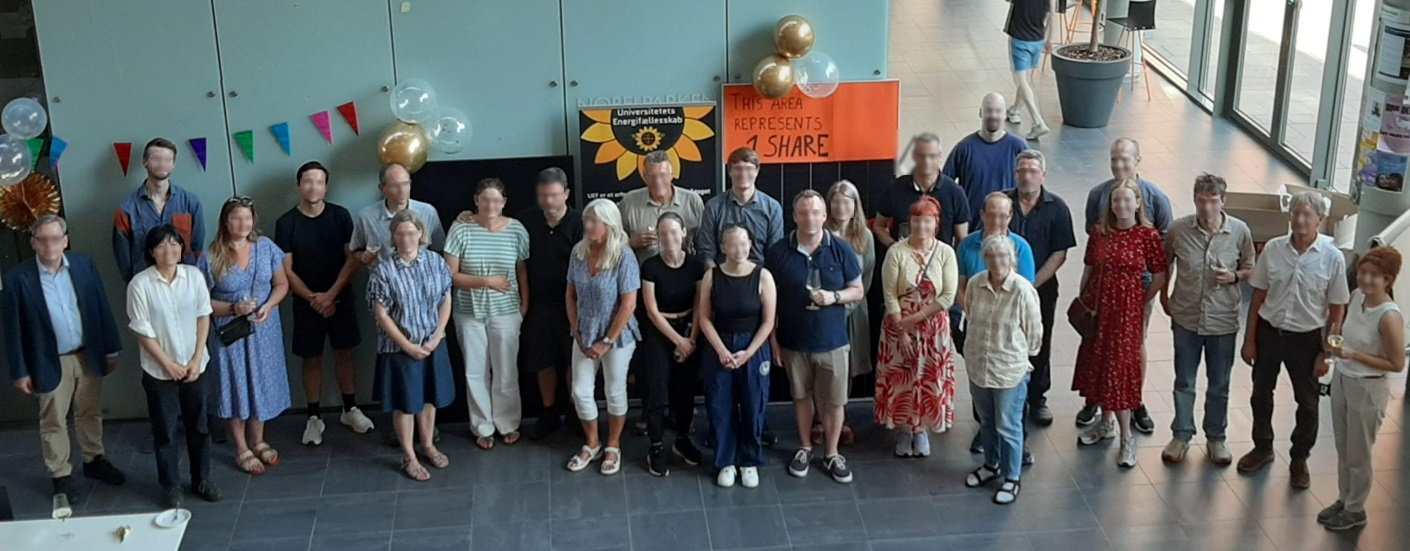}
\caption{Photograph of some of the energy community participants at the inauguration party in September 2024.}  
\end{figure}

\begin{figure}[!h]
\centering
\includegraphics[width=0.4\linewidth]{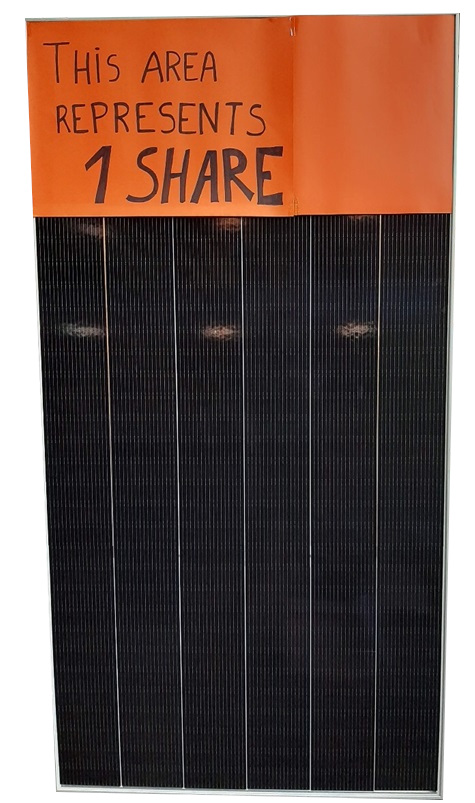}
\caption{Area of the PV module corresponding to one of the 900 shares in which the energy community is divided} \label{} 
\end{figure}

\section{Putting our challenges into perspective. Similarities with existing literature}

Many researchers have previously attempted to summarize the challenges and barriers associated with implementing energy communities. These challenges span across regulatory, financial, technical, and organizational dimensions. In this section, we summarize previous literature and reflect on where it aligns or contradicts our experience. 

\

Across several studies in Europe, regulatory and policy barriers are frequently cited as the top obstacle to energy community implementation \cite{Brummer_2018, Sebi_2020, Lazdins_2021, Yiasoumas_2023, Dioba_2024, Aoidh}. In some cases, the issue is the lack of clear regulatory and legal framework supporting the creation and operation of energy communities \cite{Brummer_2018, Sebi_2020, Lazdins_2021, Yiasoumas_2023, Dioba_2024, Aoidh}. Dioba et al. \cite{Dioba_2024}, however, emphasize that the complexity of regulation, more so than legal limitations, is the most significant barrier across European Union countries. In our case, it took us some effort and required help from an external lawyer to set up the legal framework for the energy community confirming previous literature. 

\

Financial and economic constraints also play a critical role. Factors such as high capital cost and limited access to financing mechanisms are commonly identified in the literature \cite{Brummer_2018, Sebi_2020, Yiasoumas_2023, Dioba_2024, Aoidh}. In particular, several studies highlight the difficulty in securing bank loans for community energy projects\cite{Brummer_2018, Aoidh}. Coupled with uncertain or slow returns on investment \cite{Lazdins_2021, Yiasoumas_2023} and market uncertainties \cite{Dioba_2024}, these financial and economic constraints can significantly limit the implementation of energy community projects. In our case, we limited the value of one share to 900 DKK (120 EUR) to facilitate funding and participation. The drawback is the higher administrative costs to manage a cooperative composed of many members with low investment per share. The market uncertainty identified in previous literature has been clearly a large challenge in our case, as for example, the risk of negative prices has prevented the cooperative from establishing a contract to export electricity to the grid. 

\

A few studies identify the reliance on volunteers as a challenge for energy communities \cite{Brummer_2018, Sebi_2020, Wright_2023,
Herbes_2017}. This can lead to both organizational and technical challenges. From an administrative perspective, volunteers typically have limited time to dedicate to projects, resulting in constrained human resources \cite{Sebi_2020, Wright_2023}. Technically, this translates to a lack of expertise and competencies \cite{Wright_2023}. As a result, communities may need to hire external experts, which in turn raises project costs and adds to the financial challenge \cite{Brummer_2018}. Here, we have benefited from European funding, via the AURORA project, to hire a community coordinator, as well as benefit from large voluntary work from members. The hiring of a community coordinator was key to kickstart the project. 

\

Additionally, both Aoidh et al. \cite{Aoidh} and Dioba et al. \cite{Dioba_2024} , point to low trust and mistrust in the community model as further challenges. A few sources cite that lack of information, whether technical, legal, or financial, as a hurdle \cite{Cristobal_2023,
De_franco_2023}. However, De Franco et al  technical, legal, or financial, as a hurdle \cite{Cristobal_2023} argue that it is not necessarily the lack of comprehendible information but rather how information is interchanged and communicated that challenges community engagement. In our experience, this has not been a barrier, but the community members showed a high level of trust in the board of the energy community. The plans and actions taken were communicated through a mailing group and general assemblies are held annually.

\section{Conclusions and policy recommendations}

In this manuscript, we have described the main challenges that we faced when setting up an energy community participated by students and employees at Aarhus University in Denmark. We hope that our description could be inspiring and helpful for others trying to attempt to create an energy community. We would also like to share some ideas and policy recommendations at different levels to ease the deployment of energy communities. 

\

At the European level, it would be desirable that the legal definition of shareholders in the European Directive on the promotion of the use of energy from renewable sources
\cite{definition} includes universities, hospitals, administrative buildings and other institutions, that although not administrated at the municipal level, have the potential to become energy communities. 

\

At the national energy regulation in Denmark, policy recommendations include (i) establishing legally the possibility for universities to become energy communities, (ii) removing the electricity tax (`\textit{elafgift}') on the locally-produced electricity under the scheme of self-consumption via third party when the owner of the installation or the consumer of the electricity is an energy community (Table \ref{tab_components}), and (iii) relieving the energy communities of the need to pay negative electricity prices when they export electricity to the grid. On top of all the social and environmental benefits brought by energy communities,  the second proposal is an extension of the electricity tax exemption currently only valid for normal consumers under self-consumption regulation.  The third suggestion would improve the economic viability of energy communities by reducing their associated economic risk at a very low cost for society. 

\

At the municipal level, energy communities can be fostered by (i) creating the figure of energy-community-enabler public employees as described in Section \ref{sec_replicating}, (ii) guaranteeing easy access to building data for a fast evaluation of static load for new rooftop PV installation, (iii) creating an easy-to-use platform to crowdsource the fundings for energy communities.  

\

Energy communities can be the strategy to enable collective and citizen-driven energy actions to support the clean energy transition, but for this to be true, it is needed to simplify the administrative work, set up reasonable economic frameworks and collect the know-how and experience through online repositories and energy-community-enabler public employees.

\section{Acknowledgements}
M.V., Z.Z., G.B.A., and P.R are partially funded by the AURORA project supported by the European Union’s Horizon 2020 research and innovation programme under grant agreement No. 101036418. We acknowledge the voluntary participation of the energy community board members, as well as several employees at Aarhus University who helped us make this project possible. We are also very grateful for the help from our legal advisor Erik Christiansen. We thank our colleagues at the Technical University of Madrid (Spain), University of Evora (Portugal), Ljubljana University (Slovenia) and Forest of Dean District Council (UK) for the many discussions and feedback on how to implement energy communities.


\begin{thebibliography}{10}
\expandafter\ifx\csname url\endcsname\relax
  \def\url#1{\texttt{#1}}\fi
\expandafter\ifx\csname urlprefix\endcsname\relax\def\urlprefix{URL }\fi
\providecommand{\bibinfo}[2]{#2}
\providecommand{\eprint}[2][]{\url{#2}}

\bibitem{Rahdan_2024}
\bibinfo{author}{Rahdan, P.}, \bibinfo{author}{Zeyen, E.},
  \bibinfo{author}{Gallego-Castillo, C.} \& \bibinfo{author}{Victoria, M.}
\newblock \bibinfo{title}{Distributed photovoltaics provides key benefits for a
  highly renewable {European} energy system}.
\newblock \emph{\bibinfo{journal}{Applied Energy}}
  \textbf{\bibinfo{volume}{360}}, \bibinfo{pages}{122721}
  (\bibinfo{year}{2024}).
\newblock
  \urlprefix\url{https://www.sciencedirect.com/science/article/pii/S0306261924001041}.

\bibitem{Clean_Planet}
\bibinfo{title}{A {Clean} {Planet} for all - {A} {European} strategic long-term
  vision for a prosperous, modern, competitive and climate neutral economy.
  {In}-depth analysis accompanying the {Communication}}.
\newblock \bibinfo{type}{Tech. Rep.} (\bibinfo{year}{2018}).
\newblock
  \urlprefix\url{https://ec.europa.eu/clima/sites/clima/files/docs/pages/com_2018_733_analysis_in_support_en_0.pdf}.

\bibitem{Oshaughnessy_2021a}
\bibinfo{author}{O’Shaughnessy, E.}
\newblock \bibinfo{title}{Toward a more productive discourse on rooftop solar
  and energy justice}.
\newblock \emph{\bibinfo{journal}{Joule}} \textbf{\bibinfo{volume}{5}},
  \bibinfo{pages}{2535--2539} (\bibinfo{year}{2021}).
\newblock
  \urlprefix\url{https://www.sciencedirect.com/science/article/pii/S2542435121003901}.

\bibitem{Oshaughnessy_2021b}
\bibinfo{author}{O’Shaughnessy, E.}, \bibinfo{author}{Barbose, G.},
  \bibinfo{author}{Wiser, R.}, \bibinfo{author}{Forrester, S.} \&
  \bibinfo{author}{Darghouth, N.}
\newblock \bibinfo{title}{The impact of policies and business models on income
  equity in rooftop solar adoption}.
\newblock \emph{\bibinfo{journal}{Nature Energy}} \textbf{\bibinfo{volume}{6}},
  \bibinfo{pages}{84--91} (\bibinfo{year}{2021}).
\newblock \urlprefix\url{https://www.nature.com/articles/s41560-020-00724-2}.
\newblock \bibinfo{note}{Number: 1 Publisher: Nature Publishing Group}.

\bibitem{Fox_2023}
\bibinfo{author}{Fox, N.}
\newblock \bibinfo{title}{Increasing solar entitlement and decreasing energy
  vulnerability in a low-income community by adopting the {Prosuming}
  {Project}}.
\newblock \emph{\bibinfo{journal}{Nature Energy}} \textbf{\bibinfo{volume}{8}},
  \bibinfo{pages}{74--83} (\bibinfo{year}{2023}).
\newblock \urlprefix\url{https://www.nature.com/articles/s41560-022-01169-5}.
\newblock \bibinfo{note}{Publisher: Nature Publishing Group}.

\bibitem{AURORA_webpage}
\bibinfo{title}{{AURORA} project webpage, accessed 05/05/2024}.
\newblock \urlprefix\url{https://www.aurora-h2020.eu}.

\bibitem{definition}
\bibinfo{title}{Directive {(EU)} 2018/2001 of the {European} {Parliment} and
  the {Council} of 11 december 2018 on the promotion of the use of energy from
  renewable sources}.
\newblock \bibinfo{type}{Tech. Rep.} (\bibinfo{year}{2018}).
\newblock \urlprefix\url{https://eur-lex.europa.eu/eli/dir/2018/2001/oj}.

\bibitem{self_consumption_DK}
\bibinfo{title}{Vejledning om reglerne om egetforbrug af elektricitet fra
  kommunale og regionale solcelleanlæg, accessed 05/05/2024}.
\newblock
  \urlprefix\url{https://www.retsinformation.dk/eli/retsinfo/2022/9979}.

\bibitem{statues}
\bibinfo{title}{Statutes of the cooperative}.
\newblock \urlprefix\url{https://www.uef.dk/about-us}.

\bibitem{DEA}
\bibinfo{title}{Technology catalogue, danish energy agency, accessed
  05/05/2024}.
\newblock
  \urlprefix\url{https://ens.dk/en/our-services/technology-catalogues/technology-data-generation-electricity-and-district-heating}.

\bibitem{Brummer_2018}
\bibinfo{author}{Brummer, V.}
\newblock \bibinfo{title}{Community energy – benefits and barriers: {A}
  comparative literature review of {Community} {Energy} in the {UK}, {Germany}
  and the {USA}, the benefits it provides for society and the barriers it
  faces}.
\newblock \emph{\bibinfo{journal}{Renewable and Sustainable Energy Reviews}}
  \textbf{\bibinfo{volume}{94}}, \bibinfo{pages}{187--196}
  (\bibinfo{year}{2018}).
\newblock
  \urlprefix\url{https://www.sciencedirect.com/science/article/pii/S1364032118304507}.

\bibitem{Sebi_2020}
\bibinfo{author}{Sebi, C.} \& \bibinfo{author}{Vernay, A.-L.}
\newblock \bibinfo{title}{Community renewable energy in {France}: {The} state
  of development and the way forward}.
\newblock \emph{\bibinfo{journal}{Energy Policy}}
  \textbf{\bibinfo{volume}{147}}, \bibinfo{pages}{111874}
  (\bibinfo{year}{2020}).
\newblock
  \urlprefix\url{https://www.sciencedirect.com/science/article/pii/S0301421520305905}.

\bibitem{Lazdins_2021}
\bibinfo{author}{Lazdins, R.}, \bibinfo{author}{Mutule, A.} \&
  \bibinfo{author}{Zalostiba, D.}
\newblock \bibinfo{title}{{PV} {Energy} {Communities}—{Challenges} and
  {Barriers} from a {Consumer} {Perspective}: {A} {Literature} {Review}}.
\newblock \emph{\bibinfo{journal}{Energies}} \textbf{\bibinfo{volume}{14}},
  \bibinfo{pages}{4873} (\bibinfo{year}{2021}).
\newblock \urlprefix\url{https://www.mdpi.com/1996-1073/14/16/4873}.

\bibitem{Yiasoumas_2023}
\bibinfo{author}{Yiasoumas, G.} \emph{et~al.}
\newblock \bibinfo{title}{Key {Aspects} and {Challenges} in the
  {Implementation} of {Energy} {Communities}}.
\newblock \emph{\bibinfo{journal}{Energies}} \textbf{\bibinfo{volume}{16}},
  \bibinfo{pages}{4703} (\bibinfo{year}{2023}).
\newblock \urlprefix\url{https://www.mdpi.com/1996-1073/16/12/4703}.

\bibitem{Dioba_2024}
\bibinfo{author}{Dioba, A.} \emph{et~al.}
\newblock \bibinfo{title}{Identifying key barriers to joining an energy
  community using {AHP}}.
\newblock \emph{\bibinfo{journal}{Energy}} \textbf{\bibinfo{volume}{299}},
  \bibinfo{pages}{131478} (\bibinfo{year}{2024}).
\newblock
  \urlprefix\url{https://www.sciencedirect.com/science/article/pii/S0360544224012519}.

\bibitem{Aoidh}
\bibinfo{title}{{A. N. Aoidh et al. }, {PESTLE Analysis} of barriers to
  community energy development, accessed 05/05/2024}.
\newblock
  \urlprefix\url{https://leco.interreg-npa.eu/subsites/leco/PESTLE_Analysis_LECO_A4_190110-singlepages.pdf}.

\bibitem{Wright_2023}
\bibinfo{author}{Wright, C.~G.}
\newblock \bibinfo{title}{Chapter 22 - {Energy} democracy cooperatives:
  {Opportunities} and challenges}.
\newblock In \bibinfo{editor}{Nadesan, M.}, \bibinfo{editor}{Pasqualetti,
  M.~J.} \& \bibinfo{editor}{Keahey, J.} (eds.)
  \emph{\bibinfo{booktitle}{Energy {Democracies} for {Sustainable} {Futures}}},
  \bibinfo{pages}{195--204} (\bibinfo{publisher}{Academic Press},
  \bibinfo{year}{2023}).
\newblock
  \urlprefix\url{https://www.sciencedirect.com/science/article/pii/B978012822796100022X}.

\bibitem{Herbes_2017}
\bibinfo{author}{Herbes, C.}, \bibinfo{author}{Brummer, V.},
  \bibinfo{author}{Rognli, J.}, \bibinfo{author}{Blazejewski, S.} \&
  \bibinfo{author}{Gericke, N.}
\newblock \bibinfo{title}{Responding to policy change: {New} business models
  for renewable energy cooperatives – {Barriers} perceived by cooperatives’
  members}.
\newblock \emph{\bibinfo{journal}{Energy Policy}}
  \textbf{\bibinfo{volume}{109}}, \bibinfo{pages}{82--95}
  (\bibinfo{year}{2017}).
\newblock
  \urlprefix\url{https://www.sciencedirect.com/science/article/pii/S0301421517304056}.

\bibitem{Cristobal_2023}
\bibinfo{author}{Cristóbal, A.~B.} \emph{et~al.}
\newblock \bibinfo{title}{Delving into the modeling and operation of energy
  communities as epicenters for systemic transformations}.
\newblock \emph{\bibinfo{journal}{Universal Access in the Information Society}}
   (\bibinfo{year}{2023}).
\newblock \urlprefix\url{https://doi.org/10.1007/s10209-023-01056-0}.

\bibitem{De_franco_2023}
\bibinfo{author}{De~Franco, A.} \emph{et~al.}
\newblock \bibinfo{title}{Drivers, {Motivations}, and {Barriers} in the
  {Creation} of {Energy} {Communities}: {Insights} from the {City} of
  {Segrate}, {Italy}}.
\newblock \emph{\bibinfo{journal}{Energies}} \textbf{\bibinfo{volume}{16}},
  \bibinfo{pages}{5872} (\bibinfo{year}{2023}).
\newblock \urlprefix\url{https://www.mdpi.com/1996-1073/16/16/5872}.

\end{thebibliography}
\end{document}